# Validation: Conceptual versus Activity Diagram Approaches

Sabah Al-Fedaghi

Computer Engineering Department
Kuwait University
Kuwait

*Abstract*—A conceptual model is used to support development and design within the area of systems and software modeling. The notion of validation refers to representing a domain in a model accurately and generating results using an executable model. In UML specifications, validation verifies the correctness of UML diagrams against any constraints and rules defined within the model. Currently, significant research has been conducted on generating test sets to validate that UML diagrams conform to requirements. UML activity diagrams are a specific focus of such efforts. An activity diagram is a flexible instrument for describing a system's behaviors and the internal logic of complex operations. This paper focuses on the notion of validation using activity diagrams and contrasts that process with a proposed method that involves an informal validation procedure. Accordingly, this informal validation involves comparing requirements to specifications expressed by a diagram of a modeling language called thinging machine (TM) modeling. The informal validation is a type of model checking that requires the model to be small enough for the verification to be done in a limited space or time period. In the proposed method, the model diagram is divided into subdiagrams to achieve this purpose. We claim the TM behavioral model comes with a particular dispositional structure that allows a designer to "carve" a model into smaller components for informal validation, which is shown through two case studies.

## I. INTRODUCTION

A conceptual model is a mathematical/logical/verbal representation (mimic) of a domain (real or proposed), situation, policy, or phenomenon developed for a particular study [1][2][3]. A conceptual (in contrast to an intentional mental representation such as sensation [4]) model describes "how we conceive of that domain" [5]. It is used to support development and design within the area of systems and software modeling (e.g., databases or business processes).

An example of such a model is a description developed using the Unified Modeling Language (UML) to construct a representation of a domain using primitive constructs and concepts such as the "lens" through which reality is perceived to capture that domain's meaning [5]. Nonconceptual models such as mathematical models are presented in terms of variables and quantitative relationships (i.e., equations). By contrast, in conceptual models, the variables and relationships between variables are represented visually as a system of icons in a diagram [5]. Visual representations can help to shift the focus to enhanced qualitative conceptual reasoning, serve as representations of an internal (mental) model, and provide a means for communication and analysis. In the context of conceptual modeling, to ensure a system's quality, it is critical that the model that represents the domain be semantically correct. This is confirmed by checking that the model satisfies some correctness properties and the system requirements.

### A. Validation

Validation is the process of confirming that models are understood, defined well, documented, and based on established fundamentals [6][7]. It conveys a sense that "a scientific effort must be justified in some logical, objective, and algorithmic way" [6]. However, determining whether a particular model fulfills requirements by validating it over the complete domain of its intended applicability often is not possible (e.g., due to cost and time). Instead, tests and logical reasoning are conducted until adequate assurance is achieved that the model can be considered valid for its intended application.

Regarding UML, as a semi-informal notation, significant research efforts have gone into the so-called model-driven testing of UML diagrams. Such efforts mostly involve generating high-level test cases that can be used to validate both specifications and implementations [8]. Specifically, activity diagrams are highly useful for validating requirements with customer representatives [9]. Activity diagrams have become an established modeling notation for various levels of abstraction, ranging from fine-grained descriptions of algorithms to high-level workflow models in business applications [10].

### B. Problem: Semantics of the UML Diagram

According to Tariq, Sang, Gulzar, and Xiang [11], the absence of formal semantics for UML activity diagrams makes it difficult to build automated tools for analyzing and validating such diagrams. Recently, UML 2.0 introduced token-driven semantics for activity diagrams inspired by Petri nets. One of the goals of the Foundational UML Subset (fUML [12]) is to provide a well-defined execution of UML activity diagrams. Accordingly, additional validation tools are needed for diagrammatic representation in the context of conceptual modeling. This paper proposes such a tool using a new type of conceptual model.

### C. Approach and Limitations

This paper focuses on works that solely examine the notion of validation using activity diagrams, as an example of the current state of research in the validation field. Validation of UML activity diagrams using directed test cases is very promising [8]. The present paper complements such studies by examining validation under "equivalent" representations using a diagrammatic model based on thinging machine (TM) modeling. The aim is to propose a particular technique for model validation based on TM.

## II. REVIEW

An UML activity diagram is a semi-formal semantic specification that is intuitive and flexible. It is used to describe a system's behaviors and the internal logic of complex operations. Therefore, it is widely utilized as a front-end tool for system-level design of software and/or hardware systems.

### A. Review: Graphs, Petri Nets, and Event B

The validation literature on developing test cases for activity diagrams is extensive (e.g., [8]). According to Chen and Mishra [8], "Most directed test case generation work is performed by human intervention. Hand-written test cases entail laborious and time-consuming effort of verification engineers who have deep knowledge of the design. Due to the manual development, it is difficult to generate all directed test cases to achieve a coverage goal. The problem is further aggravated due to the lack of comprehensive functional coverage metrics." Many tools and methods have been developed to support test specifications and test case generation. For example, dSPACE developed a tool that uses activity diagrams for test descriptions and test script generation [13]. Chen, Poon, Tang, and Tse [14] presented a framework with which to construct test cases from specifications by identifying a set of input categories for the activity diagrams as test cases. Hettab, Kerkouche, and Chaoui [15] converted activity diagramming into grammar rules for graphs to capture all the relevant features for test case generation. Shirole, Kommuri, and Kumar [16] transformed activity diagrams into extended control flow graphs. Sunitha [17] incorporated Object Constraint Language (OCL) into activity diagramming for test case generation involving difficulties identifying complete behavior and static changes [18]. Chen et al. [19] matched Java program traces with behavior activity diagramming to identify changes resulting in a failure to identify static changes [18]. Sapna and Mohanty [20] converted UML activity diagrams into tree structures to prioritize scenarios by assigning weights to nodes and edges; however, this approach lacks in-depth code coverage and cannot identify static changes [18]. Some authors have developed frameworks to transform a UML activity diagram into Petri nets automatically using a model checker for analysis (e.g., [21][22]). A different approach involves transforming UML activity diagrams into Event B to specify and verify the distributed and parallel workflow solicitations [23].

### B. Approach in this Paper: Informal Validation

In our approach to validation, we first produce an equivalent TM representation of the activity diagram. We consider the complexity of the representation and thus aim for quick model checking by adopting informal reasoning. All of the reviewed validation methods discussed in the previous subsection can be applied to TM. Informal validation leads to discussing formal validation.

Formal specifications can be used to deliver a precise addition to different descriptions and can be validated, leading to specification faults being detected. Formal validation verifies the correctness of specifications, so it can be used to guarantee the quality of models (e.g., UML [8]). Despite the long interest in formal validation methods, "It seems that practitioners judge formal methods to be insufficiently beneficial to outweigh pragmatic problems" [24]. According to Amey [25], "Customers are often 'aghast' at the idea of formal methods being used to develop their products and might say 'couldn't you use UML?'" Amey [25] suggests overcoming such prejudices through "formality by stealth" and cites semantically strengthened UML as an example [24].

On the other hand, informal validation techniques rely on the opinions of modelers to draw a conclusion [26]. According to Petty [27], "Informal methods are more qualitative than quantitative and generally rely heavily on subjective human evaluation, rather than detailed mathematical analysis. Experts examine an artifact, for example, a conceptual model expressed as UML diagrams, and assess the model based on that examination and their reasoning and expertise." Examples of informal methods include inspection, face validation, the Turing test, desk checking, and walkthroughs [27]. According to Banks [26], "In all cases though it is important to note that informal does not mean it is any less of a true testing method. These methods should be performed with the same discipline and structure that one would expect in 'formal' methods. When executed in such a way, solid conclusions can be made."

The purpose of informal validation is to examine the accuracy of a domain's representation in a conceptual model and in the results produced by the executable model [27]. In this type of validation, the concept of a system is viewed as a group of interacting components, and its desired functionality is articulated by graphical means [28].

Accordingly, in our proposed approach, the validation process involves requirements (e.g., expressed in English) versus specifications (expressed by TM diagrams). The validation here involves showing that the TM model is the correct model for the requirements. Thus, informal validation is a type of model checking that requires "the model to be small enough so that the verification can be done in a limited space or time" [29]. Accordingly, the TM diagram is divided into subdiagrams for this purpose. Our claim is that the TM behavioral model comes with a particular dispositional structure that allows a designer to "carve" a diagram into smaller components for informal validation.

### C. Outline

The next section reviews the basic constructs of a TM model. Section 4 presents a case study of validating the process of buying a beverage from a vending machine. Section 5 presents a second case study of validating an online shopping system.

## III. THE THINGING MACHINE

TM modeling is a way of understanding how things and processes have come to be structured (see [30] and its TM-related references by the author of the present paper). As shown in Fig. 1, a TM can be described as the following generic (basic) actions:

Arrive: A thing moves to a machine.

Accept: A thing enters the machine. For simplification, we assume that all arriving things are accepted; hence, we can combine the arrive and accept stages into one stage: the receive stage.

Release: A thing is ready for transfer outside the machine.

Process: A thing is changed, but no new thing results.

Create: A new thing is born in the machine.

Transfer: A thing is input into or output from a machine.

Additionally, the TM model includes the mechanism of triggering (denoted by a dashed arrow in this study's figures), which initiates a flow from one machine to another. Multiple machines can interact with each other through the movement of things or triggering. Triggering is a transformation from one series of movements to another.

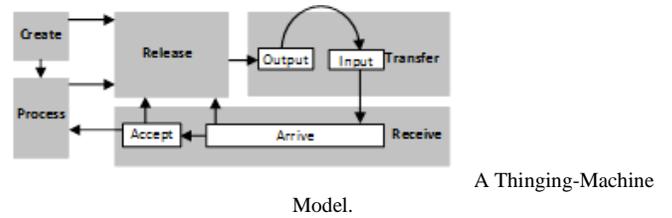

A Thinging-Machine Model.

## IV. VALIDATING A VENDING MACHINE

Sapna and Arunkumar [20] considered an example of an activity diagram for the process of buying a beverage from a vending machine (Fig. 2).

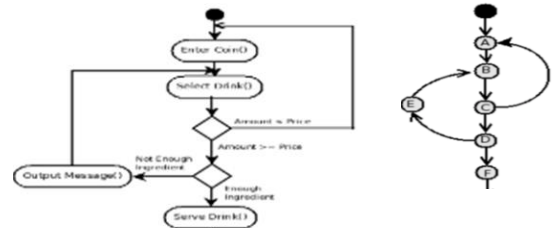

Fig. 2. Partial Views of the Diagrams used to Model the Process of Buying a Vending Machine Beverage Found in Sapna and Arunkumar [20].

In this section, we first produce the corresponding TM models and then apply the validation strategy to this vending-machine example.

### A. Static Model

Fig. 3 shows the corresponding static TM model. In Fig. 3,

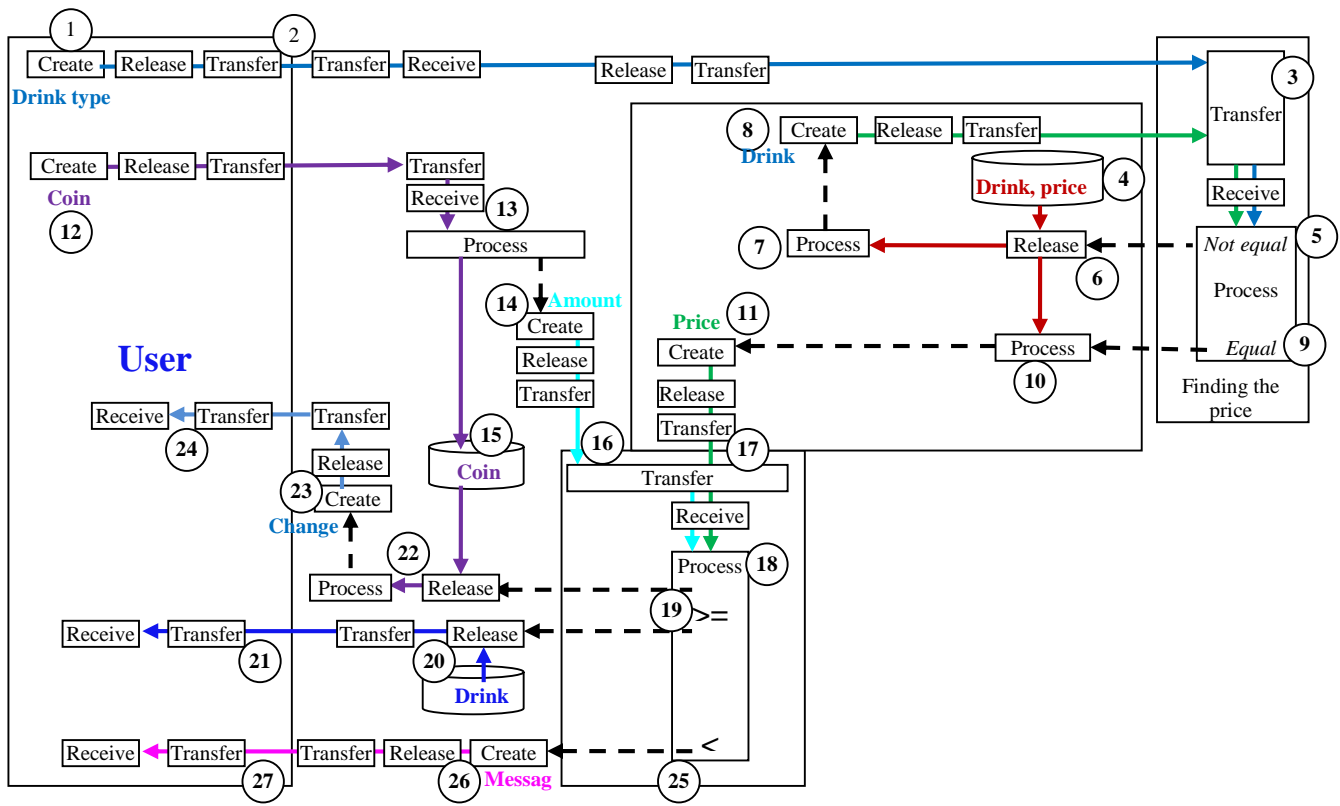

Fig. 3. The Static TM Model of a Vending Machine.

- The user creates (circle 1) a drink selection that flows to the machine (2), where it is sent to a module that finds the (drink, price) record (3). The drink data are found by extracting records (drink, price) (4) one by one and comparing the input drink with the drink data in the matching record.

  - If the drink data inside the record are not equal to the input drink (5), then the next record is released from the set of records and processed (7) to trigger the creation of drink data (8). Note that the drink item is *created* in the sense that it was not known to the machine as an independent thing before it was extracted from the record.

  - If the drink data inside the record match the input drink (9), then this triggers processing of (10) the record to extract the price (11). (Note that, in the activity diagram, the price is a thing "dropped from the sky". There is no connection between the drink and its price.)

- Meanwhile, the user inputs coins (12), which are processed (13) inside the machine to calculate their value (14). (Note that the activity diagram does not distinguish the coins as physical objects from their amount and value.) Then, the coins are stored in coin boxes (15).

- Both the amount (16) and the price (17) flow to a module that compares them (18).

  - If the amount is equal to or greater than the price (19), then the drink is released to the user (20 and 21). Additionally, the coin storage is processed (22) to create change, which flows to the user (23 and 24).

  - If the amount is less than the price (25), then a message is created and flows to the user (26 and 27).

### B. Behavioral Model

To produce the TM behavioral model, we must identify all events in the vending-machine model. An event in TM modeling is formed from a subset of the static model in addition to a time subthimac. For example, Fig. 4 shows the event *the machine receives a drink selection*.

Identifying a phenomenon's behavior involves dividing it into component parts and then fitting the behaviors of these parts into a whole. Accordingly, the static model (Fig. 3) can be divided into events as shown in Fig. 5, which shows only the regions of the events for simplification. The resulting list of events is as follows.

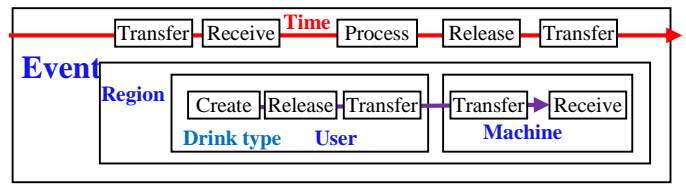

Fig. 4.    The Event the Machine Receives a Drink Selection.

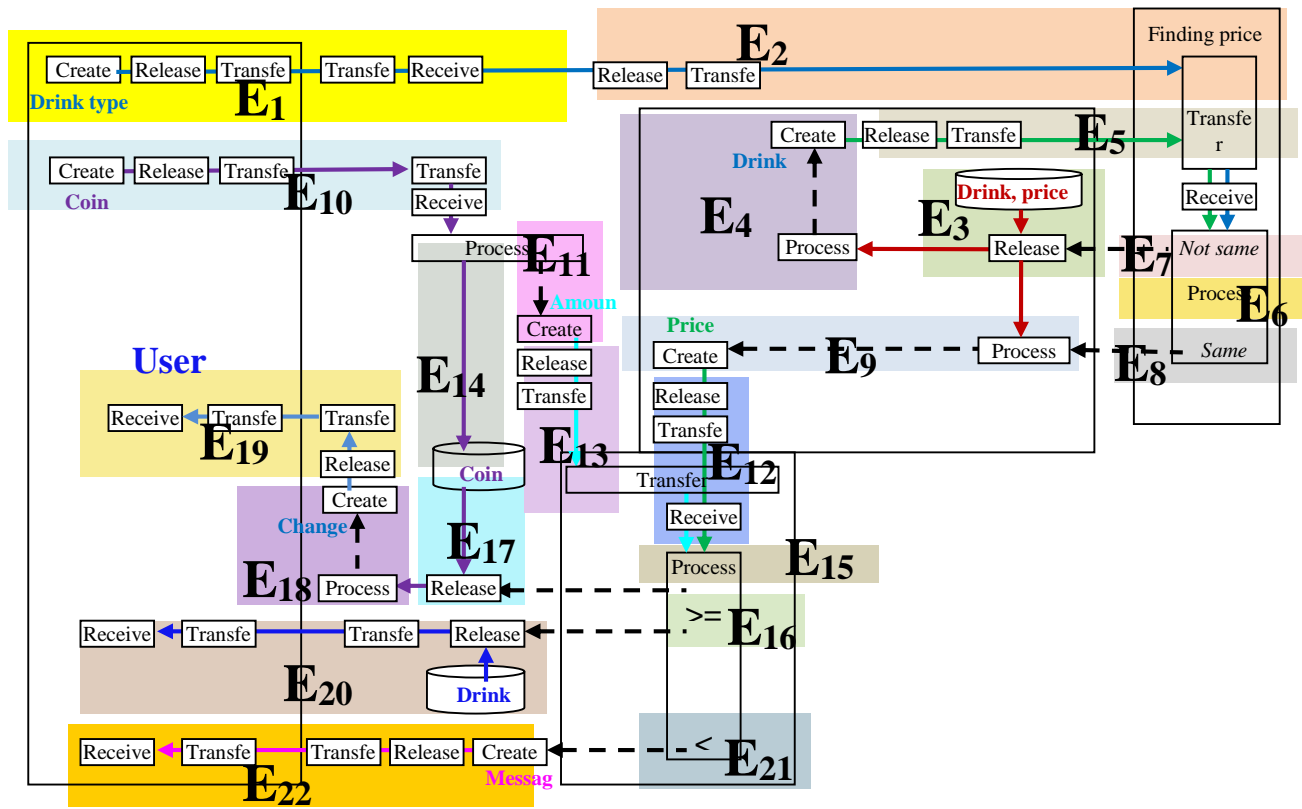

Fig. 5.    The TM Events Model of the Vending Machine.

Event 1 ($E_1$): The machine receives a drink selection.

Event 2 ($E_2$): The selected drink flows to the price-finding module.

Event 3 ($E_3$): A record (drink, prices) is retrieved from the list.

Event 4 ($E_4$): The selected drink is extracted from the record.

Event 5 ($E_5$): The drink is sent for comparison with the input drink.

Event 6 ($E_6$): The input drink is compared with the stored drink.

Event 7 ($E_7$): The input drink is not the same as the stored drink.

Event 8 ($E_8$): The input drink is the same as the stored drink.

Event 9 ($E_9$): The price is extracted.

Event 10 ($E_{10}$): The user inputs coins.

Event 11 ($E_{11}$): The amount of the coins' value is calculated.

Event 12 ($E_{12}$): The coins are deposited into the coin boxes.

Event 13 ($E_{13}$): The amount flows to a comparison with the price.

Event 14 ($E_{14}$): The price flows to a comparison with the amount.

Event 15 ($E_{15}$): The amount and the price are compared.

Event 16 ($E_{16}$): The amount is equal to or greater than the price.

Event 17 ($E_{17}$): The coin boxes are processed.

Event 18 ($E_{18}$): The change is extracted from the coin boxes.

Event 19 ($E_{19}$): The change flows to the user.

Event 20 ($E_{20}$): The drink is released to the user.

Event 21 ($E_{21}$): The input amount is less than the price.

Event 22 ($E_{22}$): A message is sent to the user.

Fig. 6 shows the behavioral model for the vending machine.

## C. Validation Strategy

Until this point, we have focused only on the modeling notations. It is time to ask whether the vending-machine blueprint fulfills the requirements. Requirements typically are written in natural language, but the behavioral diagram provides a skeletal structure of events.

Plato famously employed the "carving" metaphor as an analogy for the reality of Forms (Phaedrus 265e): As if we were animals, the world comes to us predivided. Ideally, our best theories will be those that carve nature at its joints [31]. The behavioral model comes with a particular dispositional structure rooted in the five generic types of events: create, processes, release, transfer, and receive.

Validation in TM modeling refers to event validity, which involves the model's events being compared to those of the reality to determine whether they are similar. In TM modeling, a general approach to validating the developed model can be developed using a logical process in which one takes higher-level events produced by the carving process and reduces them to the constituent events, which, in turn are based on generic events. This implies validating each generic action or sequence of these actions. However, because of space limitations and the informal nature of this study, we will not employ such a process but rather simply sketch operational descriptions of events with which to validate the model. This method assumes that the model's validity can be determined from observations.

Fig. 7 shows the decomposition of the behavioral model (Fig. 6) into three parts (super-events), for which the joints suggest division among three super-events as follows.

- Super-event 1: Selecting a drink and finding the price

  - The machine shall accept requests for *n* types of drinks.

  - The machine determines the price of the selected drink.

Fig. 8 shows the events involved: $E_1$ through $E_9$. The figure shows that the verification method involves feeding, internally, all drinks stored in the machine to the machine to verify that the machine performs the two requirements specified above. This verification process covers all legitimate inputs to $E_1$. An actual verification system can be constructed to input the tuples to $E_1$.

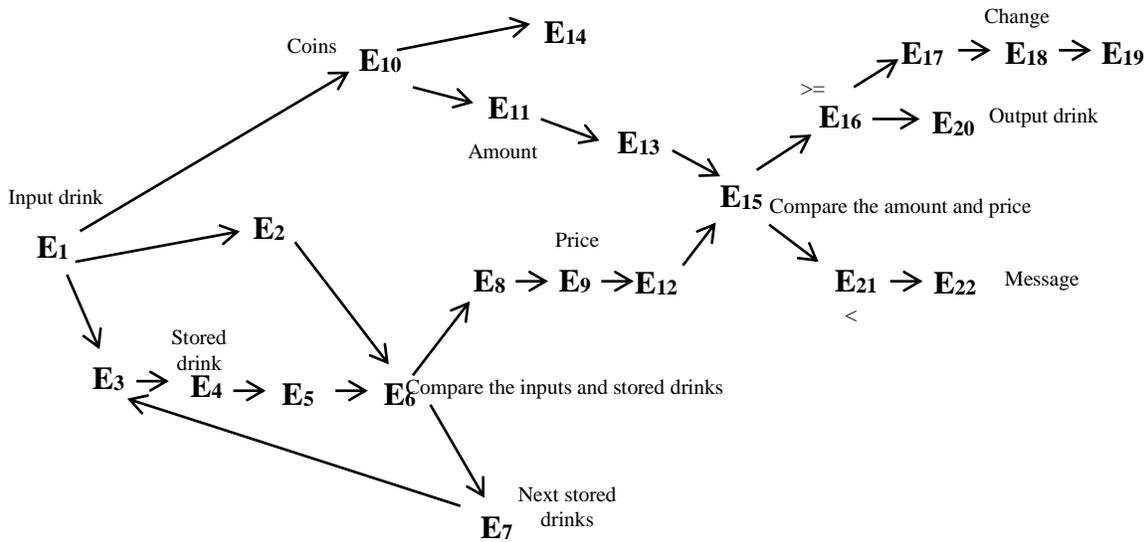

Fig. 6. The Behavioral Model.

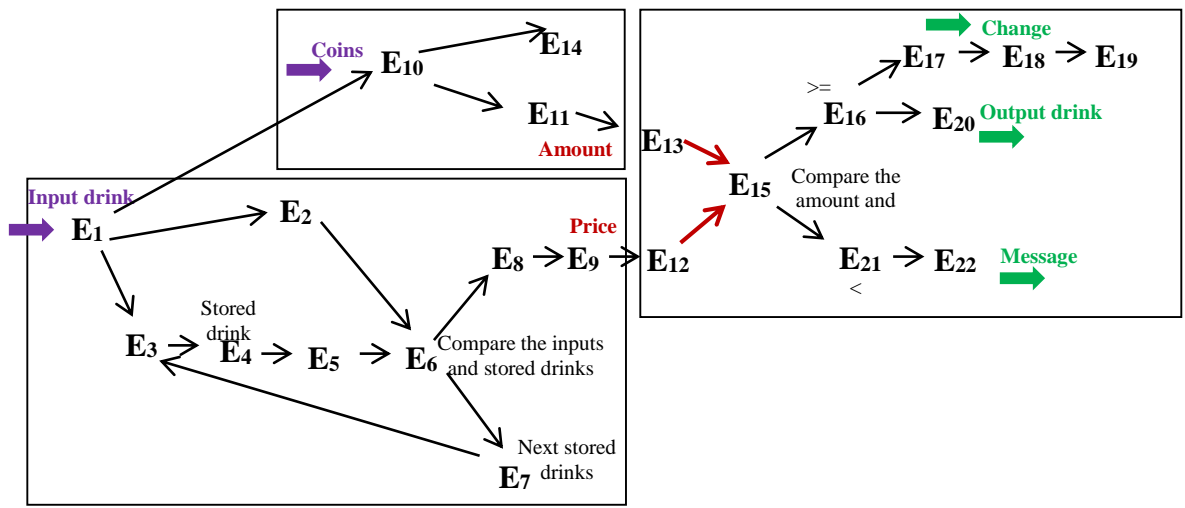

Fig. 7. Three Super-Events in the Behavioral Model.

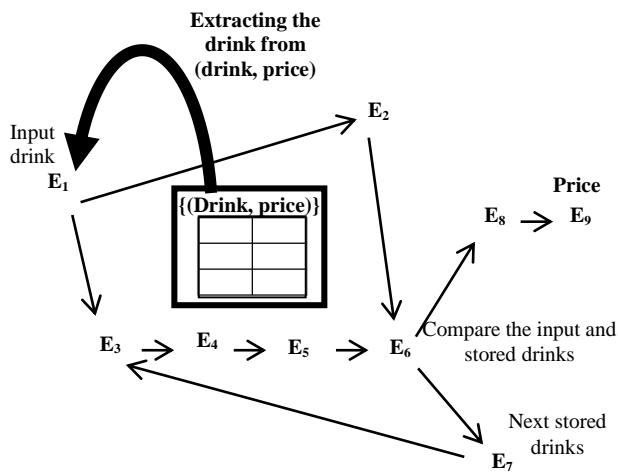

Fig. 8. Verifying that All Drinks have Prices.

*Verification of selecting a drink and finding the price*

*Verification version of $E_1$: loop for all drinks:*
*Read a tuple from the (drink, price)*
*Extract the drink from the tuple*
*Perform $E_2$, $E_3$, $E_4$, $E_5$, $E_6$, $E_7$, $E_8$, and $E_9$*
*If (ERROR), then process a report*

- upper-event 2: Validating the coins

    - The machine accepts all specified coin types, assuming there are three types.

    - The machine calculates the digital values of these coins.

    - The machine stores the coins in their appropriate places.

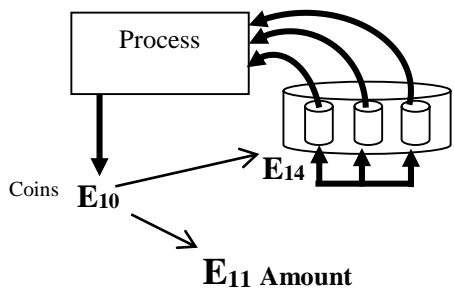

Fig. 9. Validating All Combinations of Coins to be sorted in their Boxes and Generating the Right Amount.

Fig. 9 shows the validation of all combinations of coins to be stored in their boxes and the generation of the correct amount.

We assume that the machine initially has some coins from each of the assumed three types of coins, to return change when the input amount is greater than the price. For validation purposes, this initial amount is increased to cover all types of combinations of input amounts. Accordingly, different combinations of coins are fed to $E_{10}$, which are distributed to their appropriate places ($E14$), and their amount is generated ($E_{11}$). It is not difficult to develop such an internal system in a vending-machine factory with which to test each machine.

- Super-event 3: Comparing the amount and price and outputting the result

The third validation process involves comparing the amount and price and observing the results of that comparison, as shown in Fig. 10.

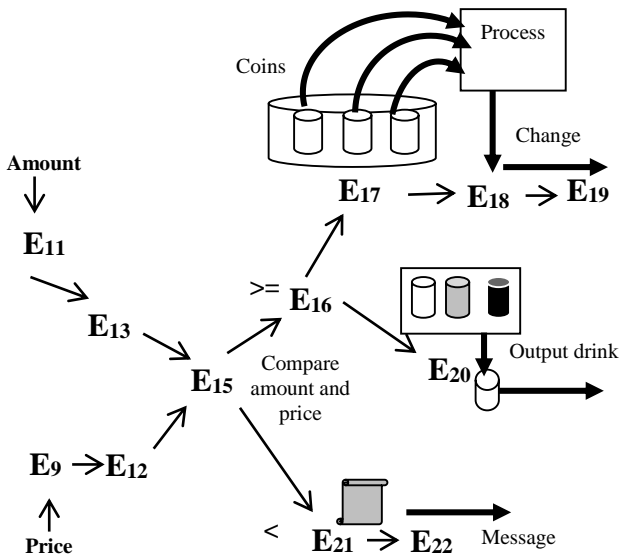

Fig. 10. Validating All Types of Output.

- All prices and amounts produced in phases 1 and 2 are fed to $E_{15}$, during which the amount and price are compared. The results of this comparison are as follows.

If the amount is less than the price, then a message is produced ($E_{21}$ and $E_{22}$).

If the amount is equal to or greater than the price, then the correct change is produced by processing the coins ($E_{17}$, $E_{18}$, and $E_{19}$).

- A drink is output.

We assume that a validation system takes the outputs of super-events 1 and 2 and produces the physical activities above in super-event 3. The level at which all possible variations of inputs are exhausted depends on the amount and price values produced in the first two super-events.

Nevertheless, in reality, the level of testing is a subjective decision based on evaluations conducted as part of the model-development procedure.

## V. VALIDATING AN ONLINE SHOPPING SYSTEM

Bures, Ahmed, and Zamli [32] proposed a model-based test-case-generation algorithm that uses directed graphs and test requirements to model the system being tested. They proposed a method using a directed graph and a set of test requirements to try to satisfy a defined test-coverage level together. They modeled an online shopping system, as presented in Fig. 11, as a running example to document the presented concepts and algorithms. Fig. 12 shows the TM model constructed to reflect the given activity diagram of Fig. 11.

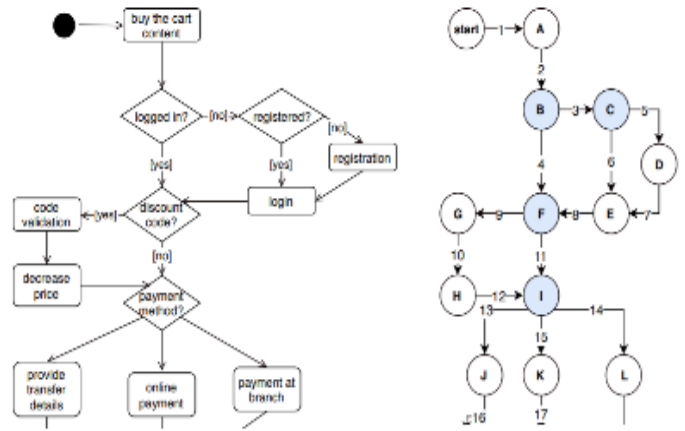

Fig. 11. Partial Views of the Diagrams used in Modeling the Online Shopping System in [31].

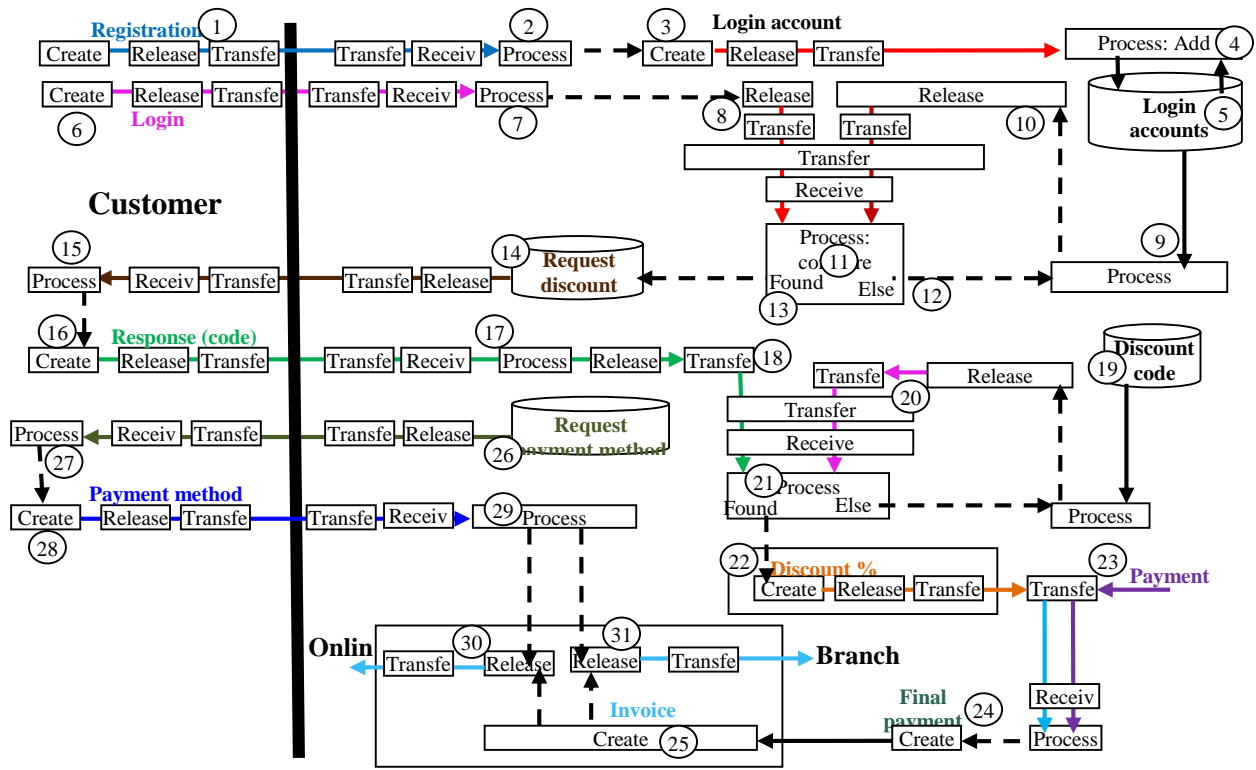

Fig. 12. The Static TM Model of the Online Shopping System.

First, the customer applies for registration (circle 1), which is processed by the system (2) to create a login account (3) that is added to the set of registered accounts (4 and 5). Note how the login account file is processed to add a new account.

- The customer requests to log in (6) and the request is processed (7) to extract the login account from the request (8). Additionally, the accounts file is processed (9) to retrieve an account (10), which is compared with the input account (11). If the two accounts are not the same, then the next account in the accounts file is retrieved for comparison. This process continues until the two accounts are found to be the same (13). Here, we ignore the situation in which the account is not filed in the file because the activity diagram does not mention it. Here, we can add a trigger for an error message to flow to the customer.

- Then, the system sends a request for a discount (14), which is processed by the user (15), to reply (16) with a code (*no discount* is a type of code). The code is processed (17) to be compared with the set of codes (18, 19, 20, and 21). When the code is found, the percentage of discount is calculated (22) and, together with the price (23), is processed to calculate the payment (24) and invoice (25).

- Accordingly, the system requests the method of payment (26), and the customer processes (27) that request to input such a method (28), which flows to the system, where it is processed (29). According to the payment method, the system either sends an invoice online (30) or to the branch (31).

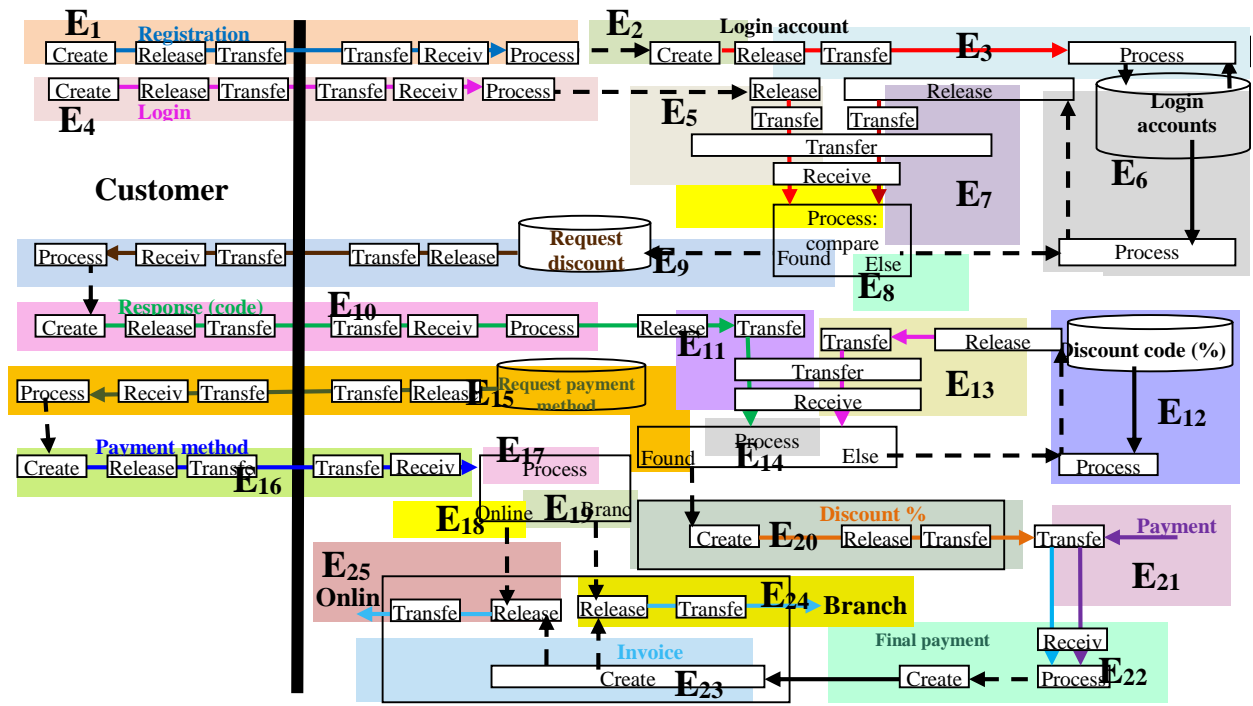

Fig. 13. Events Model. Note that in the Activity Diagram, the Price is considered an Input to the Whole Process.

The events in the model can be specified as follows (see Fig. 13).

Event 1 ($E_1$): A customer registers to log in.

Event 2 ($E_2$): The system creates a new login account.

Event 3 ($E_3$): The system adds the new account to the accounts file.

Event 4 ($E_4$): A customer sends a login request.

Event 5 ($E_5$): The system extracts the login account from the request and sends it to be checked as a legal account.

Event 6 ($E_6$): The accounts file is processed to retrieve an account, which is sent for comparison with the input account.

Event 7 ($E_7$): The input account is compared with the account retrieved from the file.

Event 8 ($E_8$): The input account is not the same as the account from the file.

Event 9 ($E_9$): The input account is found among the legitimate accounts; hence, a request for the discount code is sent to the customer.

Event 10 ($E_{10}$): The customer sends a discount code (possibly a code for no discount).

Event 11 ($E_{11}$): The code is sent to find its corresponding discount percentage.

Event 12 ($E_{12}$): The list of codes is processed to retrieve one code at a time.

Event 13 ($E_{13}$): The retrieved code is sent to be processed.

Event 14 ($E_{14}$): The code is compared with the list of codes.

Event 15 ($E_{15}$): The code is found; thus, a request for the payment method is sent to the customer.

Event 16 ($E_{16}$): The customer sends the payment method.

Event 17 ($E_{17}$): The payment method is processed.

Event 18 ($E_{18}$): The payment method is in the branch.

Event 19 ($E_{19}$): The online payment method is chosen.

Event 20 ($E_{20}$): The code is found; thus, the discount percentage is extracted.

Event 21 ($E_{21}$): The price is received.

Event 22 ($E_{22}$): The discount percentage and price are used to calculate the required payment.

Event 23 ($E_{23}$): The payment is used in generating the invoice.

Event 24 ($E_{24}$): The invoice is sent to the branch.

Event 25 ($E_{25}$): The invoice is sent to the online payment system.

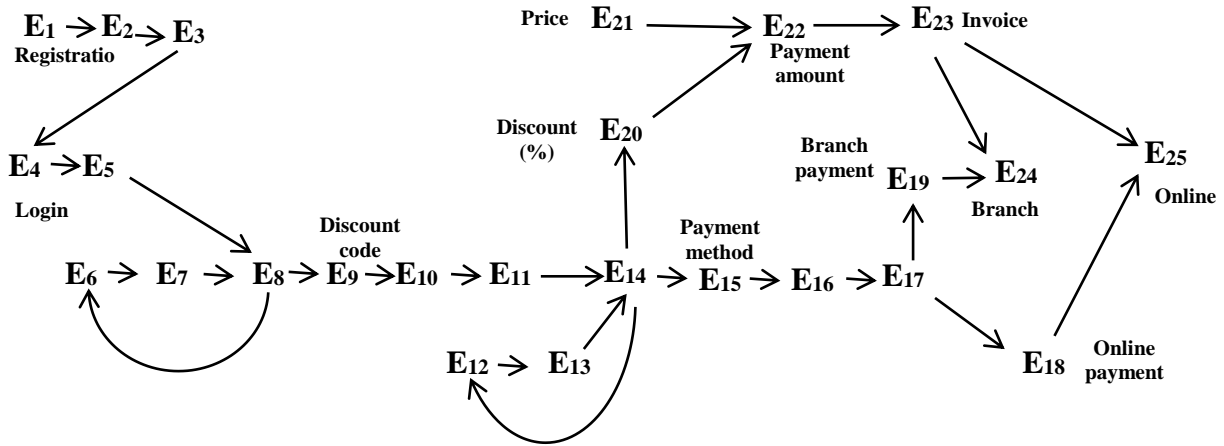

Fig. 14. The Behavioral Model.

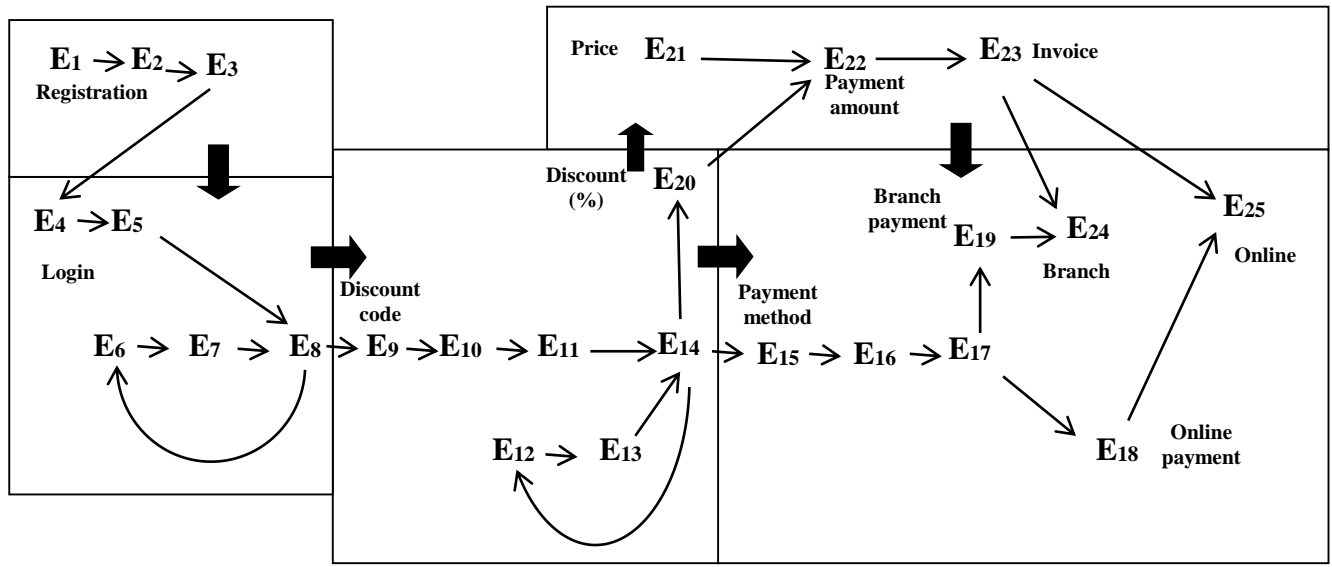

Fig. 15. Carving Small Components from the Behavioral Model.

Fig. 14 shows the behavioral model decomposed into three parts (super-events), in which the joints suggest division among five super-events. As shown in Fig. 15, these super-events are as follows.

- Registration component

- Login component

- Discount percentage component

- Payment component, and

- Method-of-payment component

Apparently, our hypothesis that TM representation would lend itself to such division of high-level events is true for this shopping-system representation. We can apply the same type of informal validation.

However, it is important to point out that the super-events may have relationships with "use cases"; thus, UML use case diagrams are a topic to be investigated in future research.

## VI. CONCLUSION

This paper focused on examining the notion of validation using activity diagrams and proposed an informal validation process. This validation process involved requirements, versus specifications expressed by a diagram. Informal validation is a type of model checking that requires the model to be small enough to verify in a limited space or time. Accordingly, the model diagram is divided into subdiagrams for this purpose. We claimed that the TM behavioral model comes with a particular dispositional structure that allows designers to "carve" a diagram into smaller components for informal validation. This was shown through two case studies concerning vending machine and online shopping systems.

This result seems plausible because TM modeling is founded upon five generic actions. Thus, generic events have no subevents, and higher-level events are built from these generic events. Generic events can happen in diverse regions, and they can reoccur. It seems the building structures from the five generic actions "collapse" into smaller structural partitions according to certain aspects such as functionality. The number (e.g., $7 \pm 2$) and nature (e.g., basic) of these actions seem to be crucial features that determine the system's overall level of complexity. Additionally, the TM model (Fig. 1) seems to generate nested hierarchies or levels with loosely coupled connections (through only transfers and triggers), which inhibit large structural complexity.

These explanations are still a type of speculation and require further research to be applied in different aspects of modeling systems beyond validation.